\documentstyle[12pt]{article}
\catcode`\@=11
\def\numberbysection{\@addtoreset{equation}{section}
        \def\theequation{\thesection.\arabic{equation}}}
\numberbysection

\begin{document}

\newlength{\lno} \lno1.5cm \newlength{\len} \len=\textwidth%
\addtolength{\len}{-\lno}

\setcounter{page}{0}

\baselineskip7mm \renewcommand{\thefootnote}{\fnsymbol{footnote}} \newpage %
\setcounter{page}{0}

\begin{titlepage}     
\vspace{0.5cm}
\begin{center}
{\Large\bf Off-Shell Bethe Ansatz Equation for osp{$(2|1)$} Gaudin Magnets}\\
\vspace{1cm}
{\large A. Lima-Santos\footnote{e-mail: dals@power.ufscar.br} \hspace{.5cm} and \hspace{.5cm} W. Utiel\footnote{e-mail: pwus@iris.ufscar.br}} \\
\vspace{1cm}
{\large \em Universidade Federal de S\~ao Carlos, Departamento de F\'{\i}sica \\
Caixa Postal 676, CEP 13569-905~~S\~ao Carlos, Brasil}\\
\end{center}
\vspace{1.2cm}

\begin{abstract}
The semi-classical limit of the algebraic Bethe Ansatz  method is used to solve the 
theory of Gaudin models. Via the off-shell method we find the
spectra
 and eigenvectors of the $N-1$ independent Gaudin Hamiltonians with
symmetry 
 $osp(2|1)$. 
We also show how the off-shell Gaudin equation solves the trigonometric Knizhnik-Zamolodchikov 
equation.
\end{abstract}
\vfill
\begin{center}
\small{\today}
\end{center}
\end{titlepage}

\baselineskip6mm

\newpage{}

\section{Introduction}

In integrable models of statistical mechanics \cite{Baxter}, an important
object is the ${\cal R}$-matrix ${\cal R}(u)$, where $u$ is the spectral
parameter. It acts on the tensor product $V^{1}\otimes V^{2}$ for a given
vector space $V$ and it is the solution of the Yang-Baxter ({\small YB})
equation 
\begin{equation}
{\cal R}_{12}(u){\cal R}_{13}(u+v){\cal R}_{23}(v)={\cal R}_{23}(v){\cal R}%
_{13}(u+v){\cal R}_{12}(u)  \label{int.1}
\end{equation}
in $V^{1}\otimes V^{2}\otimes V^{3}$, where ${\cal R}_{12}={\cal R}\otimes 
{\cal I}$, ${\cal R}_{23}={\cal I}\otimes {\cal R}$, etc. and ${\cal I}$ \
is the identity matrix. If ${\cal R}$ depends on a Planck-type parameter $%
\eta $ so that ${\cal R}(u,\eta )=1+2\eta \ r(u)+{\rm o}(\eta ^{2})$, then
the ``classical $r$-matrix'' obeys the classical {\small YB} equation 
\begin{equation}
\lbrack r_{12}(u),r_{13}(u+v)+r_{23}(v)]+[r_{13}(u+v),r_{23}(v)]=0.
\label{int.2}
\end{equation}

Nondegenerate solutions of (\ref{int.2}) in the tensor product of two copies
of simple Lie algebra {\rm g} , $r_{ij}(u)\in {\rm g}_{i}\otimes {\rm g}_{j}$
, $i,j=1,2,3$, were classified by Belavin and Drinfeld \cite{BD}.

The classical {\small YB} equation interplays with conformal field theory in
the following way: In the skew-symmetric case $r_{ji}(-u)+r_{ij}(u)=0$, it
is the compatibility condition for the system of linear differential
equations 
\begin{equation}
\kappa \frac{\partial \Psi (z_{1},...,z_{N})}{\partial z_{i}}=\sum_{j\neq
i}r_{ij}(z_{i}-z_{j})\Psi (z_{1},...,z_{N})  \label{int.3}
\end{equation}
in $N$ complex variables $z_{1},...,z_{N}$ for vector-valued functions $\Psi 
$ with values in the tensor space $V=V^{1}\otimes \cdots \otimes V^{N}$ and $%
\kappa $ is a coupling constant.

In the rational case \cite{BD}, very simple skew-symmetric solutions are
known: $r(u)={\rm C}_{2}/u$, where ${\rm C}_{2}\in {\rm g}\otimes {\rm g}$
is a symmetric invariant tensor of a finite dimensional Lie algebra ${\rm g}$
acting on a representation space $V$. Then the corresponding system of
linear differential equations (\ref{int.3}) is the Knizhnik-Zamolodchikov (%
{\small KZ}) equation for the conformal blocks of the
Wess-Zumino-Novikov-Witten ({\small WZNW}) model of conformal theory on the
sphere \cite{KZ}.

The algebraic Bethe Ansatz \cite{FT} is the powerful method in the analysis
of integrable models. Besides describing the spectra of quantum integrable
systems, the Bethe Ansatz also is used to construct exact and manageable
expressions for correlation functions \cite{KBI}. Various representations of
correlators in these models were found by Korepin \cite{KO}, using this
method.

Recently, Babujian and Flume \cite{BAF} developed a method which reveals a
link to the algebraic Bethe Ansatz for the theory of the Gaudin model. In
their method the wave vectors of the Bethe Ansatz equation for inhomogeneous
lattice model render in the semi-classical limit solutions of the {\small KZ}
equation for the case of simple Lie algebras of higher rank. More precisely,
the algebraic quantum inverse scattering method permits us write the
following equation 
\begin{equation}
t(u|z)\Phi (u_{1,\cdots ,}u_{p})=\Lambda (u,u_{1},\cdots ,u_{p}|z)\Phi
(u_{1},\cdots ,u_{p})-\sum_{\alpha =1}^{p}\frac{{\cal F}_{\alpha }\Phi
^{\alpha }}{u-u_{\alpha }}.  \label{int.4}
\end{equation}
Here $t(u|z)$ denotes the transfer matrix of the rational vertex model in an
inhomogeneous lattice acting on an $N$-fold tensor product of $SU(2)$
representation spaces. $\Phi ^{\alpha }=\Phi (u_{1},\cdots u_{\alpha
-1},u,u_{\alpha +1},...,u_{p})$. ${\cal F}_{\alpha }(u_{1},\cdots ,u_{p}|z)$
and $\Lambda (u,u_{1},\cdots ,u_{p}|z)$ are $c$ numbers. The vanishing of
the so-called unwanted terms, ${\cal F}_{\alpha }=0$, is enforced in the
usual procedure of the algebraic Bethe Ansatz by choosing the parameters $%
u_{1},...,u_{p}$. In this case the wave vector $\Phi (u_{1},\cdots ,u_{p})$
becomes an eigenvector of the transfer matrix with eigenvalue $\Lambda
(u,u_{1},\cdots ,u_{p}|z)$. If we keep all unwanted terms, i.e. ${\cal F}%
_{\alpha }\neq 0$, then the wave vector $\Phi $ in general satisfies the
equation (\ref{int.4}), named in \cite{B} as off-shell Bethe Ansatz equation
({\small OSBAE}). There is a neat relationship between the wave vector
satisfying the {\small OSBAE} (\ref{int.4}) and the vector-valued solutions
of the {\small KZ} equation (\ref{int.3}): The general vector valued
solution of the {\small KZ} equation for an arbitrary simple Lie algebra was
found by Schechtman and Varchenko \cite{SV}. It can be represented as a
multiple contour integral

\begin{equation}
\Psi (z_{1},\ldots ,z_{N})=\oint \cdots \oint {\cal X}(u_{1},...,u_{p}|z)%
\phi (u_{1},...,u_{p}|z)du_{1}\cdots du_{p}.  \label{int.5}
\end{equation}
The complex variables $z_{1},...,z_{N}$ of (\ref{int.5}) are related with
the disorder parameters of the {\small OSBAE} . The vector valued function $%
\phi (u_{1},...,u_{p}|z)$ is the semi-classical limit of the wave vector $%
\Phi (u_{1},...,u_{p}|z)$. In fact, it is the Bethe wave vector for Gaudin
magnets \cite{GA}, but off mass shell. The scalar function ${\cal X}%
(u_{1},...,u_{p}|z)$ is constructed from the semi-classical limit of the $%
\Lambda (u=z_{k};u_{1},...,u_{p}|z)$ and ${\cal F}_{\alpha }(u_{1},\cdots
,u_{p}|z)$. This representation of the $N$-point correlation function shows
a deep connection between the inhomogeneous vertex models and the {\small %
WZNW }theory.

In this work we generalize previous results applying the Babujian-Flume
ideas for $osp(2|1)$ rational solution of the graded version of the {\small %
YB} equation \cite{LI1}. It is shown that this ideas persists for the case
of semi-classical limit which corresponds to the $osp(2|1)$ trigonometric $r$%
-matrix.

The paper is organized as follows. In Section $2$ we present the algebraic
structure of the $osp(2|1)$ vertex model. Here the inhomogeneous Bethe
Ansatz is read from the homogeneous case previously known \cite{LI2}. We
also derive the Off-shell Bethe Ansatz equation for the fundamental
representation of the $osp(2|1)$ algebra. In Section $3$ , taking into
account the semi-classical limit of the results presented in the Section $2$%
, we describe the algebraic structure of the corresponding Gaudin model. In
Section $4$, data of the off-shell Gaudin equation are used to construct
solutions of the trigonometric {\small KZ} equation. In Section $5$, our
results are extended for the highest representations of the algebra $%
osp(2|1) $. Conclusions are reserved for Section $6$.

\section{Structure of the ${\bf osp(2|1)}$ Vertex Model}

We recall that the $osp(2|1)$ algebra is the simplest superalgebra and it
can be viewed as the graded version of $sl_{2}$. It has three even (bosonic)
generators $H,$\ $X^{\pm }$ generating a Lie subalgebra $sl_{2}$ and two odd
(fermionic) generators $V^{\pm }$ , whose non-vanishing commutation
relations in the Cartan-Weyl basis reads as 
\begin{eqnarray}
\lbrack H,X^{\pm }] &=&\pm X^{\pm },\quad \lbrack X^{+},X^{-}]=2H,  \nonumber
\\
\lbrack H,V^{\pm }] &=&\pm \frac{1}{2}V^{\pm },\quad \lbrack X^{\pm },V^{\mp
}]=V^{\pm },\quad \lbrack X^{\pm },V^{\pm }]=0,  \nonumber \\
\{V^{\pm },V^{\pm }\} &=&\pm \frac{1}{2}X^{\pm },\quad \{V^{+},V^{-}\}=-%
\frac{1}{2}H.  \label{str.1}
\end{eqnarray}
The quadratic Casimir operator is 
\begin{equation}
C_{2}=H^{2}+\frac{1}{2}\{X^{+},X^{-}\}+[V^{+},V^{-}],  \label{str.2}
\end{equation}
where $\{\cdot \ ,\cdot \}$ denotes the anticommutator and $[\cdot \ ,\cdot
] $ the commutator.

The irreducible finite-dimensional representations $\rho _{j}$ with the
highest weight vector are parametrized by half-integer $s=j/2$ or by the
integer $j\in N$. Their dimension is {\rm dim}$(\rho _{j})=2j+1$ and the
corresponding value of $C_{2}$ is $j(j+1)/4=s(s+1/2)$, $s=0,1/2,1,3/2,...$

The representation corresponding to $s=0$ is the trivial one-dimensional
representation. The $s\geq 1/2$ representation contains two isospin
multiplets which belong to isospin $s$ and $s-1/2$, denoted by $\left|
s,s,m\right\rangle $ and $\left| s,s-1/2,m\right\rangle $, respectively. The
first quantum number characterizes the representation and the second and
third quantum numbers give the isospin and its third component. After a
convenient normalization of the states , a given $s$-representation is
defined by 
\begin{eqnarray}
H\left| s,s,m\right\rangle &=&m\left| s,s,m\right\rangle ,  \nonumber \\
X^{\pm }\left| s,s,m\right\rangle &=&\sqrt{(s\mp m)(s\pm m+1)}\left|
s,s,m\pm 1\right\rangle ,  \nonumber \\
V^{\pm }\left| s,s,m\right\rangle &=&\pm \frac{1}{2}\sqrt{(s\mp m)}\left|
s,s-1/2,m\pm 1/2\right\rangle ,  \nonumber \\
&&  \nonumber \\
\quad H\left| s,s-1/2,m\right\rangle &=&m\left| s,s-1/2,m\right\rangle , 
\nonumber \\
X^{\pm }\left| s,s-1/2,m\right\rangle &=&\sqrt{(s-1/2\mp m)(s-1/2\pm m+1)}%
\left| s,s-1/2,m\pm 1\right\rangle ,  \nonumber \\
V^{\pm }\left| s,s-1/2,m\right\rangle &=&\pm \frac{1}{2}\sqrt{(s-1/2\pm m+1)}%
\left| s,s,m\pm 1/2\right\rangle .  \label{str.3}
\end{eqnarray}
The fundamental representation has $s=1/2$ and is given by 
\begin{eqnarray}
H &=&\frac{1}{2}\left( 
\begin{array}{lll}
1 & 0 & \ \ 0 \\ 
0 & 0 & \ \ 0 \\ 
0 & 0 & -{}1
\end{array}
\right) ,\ \ X^{+}=\left( 
\begin{array}{lll}
0 & 0 & 1 \\ 
0 & 0 & 0 \\ 
0 & 0 & 0
\end{array}
\right) ,\ X^{-}=\left( 
\begin{array}{lll}
0 & 0 & 0 \\ 
0 & 0 & 0 \\ 
1 & 0 & 0
\end{array}
\right) ,  \nonumber \\
V^{+} &=&\frac{1}{2}\left( 
\begin{array}{lll}
0 & 1 & 0 \\ 
0 & 0 & 1 \\ 
0 & 0 & 0
\end{array}
\right) ,\ V^{-}=\frac{1}{2}\left( 
\begin{array}{lll}
\ \ 0 & 0 & 0 \\ 
-1 & 0 & 0 \\ 
\ \ 0 & 1 & 0
\end{array}
\right) .  \label{str.4}
\end{eqnarray}

In (\ref{str.4}) the basis is $\left| \frac{1}{2},\frac{1}{2},\frac{1}{2}%
\right\rangle ,\left| \frac{1}{2},0,0\right\rangle ,\left| \frac{1}{2},\frac{%
1}{2},-\frac{1}{2}\right\rangle $. The first and third vectors will be
considered as even and the second as odd, {\it i.e}., our grading is {\small %
BFB}.

In the $j$-representation the odd part has the form \cite{Ku1}: 
\begin{equation}
V^{+}=\left( 
\begin{array}{ccccc}
0 & V_{j-1} & 0 & \cdots & 0 \\ 
0 & 0 & V_{j-2} & \ddots & \vdots \\ 
\vdots & \ddots & \ddots & \ddots & 0 \\ 
0 & \cdots & 0 & 0 & V_{-j} \\ 
0 & \cdots & 0 & 0 & 0
\end{array}
\right) ,\quad V^{-}=\left( 
\begin{array}{ccccc}
0 & 0 & \cdots & 0 & 0 \\ 
W_{j} & 0 & \ddots & \vdots & \vdots \\ 
0 & W_{j-1} & \ddots & 0 & 0 \\ 
\vdots & \ddots & \ddots & 0 & 0 \\ 
0 & \cdots & 0 & W_{-j+1} & 0
\end{array}
\right) ,  \label{str.5}
\end{equation}
where 
\begin{eqnarray}
(V_{j-1},V_{j-2},V_{j-3},...,V_{-j}) &=&\frac{1}{2}(\sqrt{j},\sqrt{1},\sqrt{%
j-1},\sqrt{2},...,\sqrt{1},\sqrt{j}),  \nonumber \\
(W_{j},W_{j-1},W_{j-2},...,W_{-j+1}) &=&\frac{1}{2}(-\sqrt{j},\sqrt{1},-%
\sqrt{j-1},\sqrt{2,}...,-\sqrt{1},\sqrt{j}).  \label{str.6}
\end{eqnarray}
For the even part we can see from (\ref{str.3}) that $H$ is diagonal and
always has eigenvalue $0$ due to isospin integer: 
\begin{equation}
H=\frac{1}{2}{\rm diag}(j,j-1,...,1,0,-1,...,-j).  \label{str.7}
\end{equation}
Moreover, $X^{\pm }$ are given by the $sl_{2}$ composition which results in
a clear relation with the odd part: $X^{\pm }=\pm 4(V^{\pm })^{2}$.

\subsection{Off-Shell Bethe Ansatz Equation}

Consider $V=V_{0}\oplus V_{1}$ a $Z_{2}$-graded vector space where $0$ and $%
1 $ denote the even and odd parts respectively. The components of a linear
operator $A\stackrel{s}{\otimes }B$ in the graded tensor product space $V%
\stackrel{s}{\otimes }V$ result in matrix elements of the form 
\begin{equation}
(A\stackrel{s}{\otimes }B)_{\alpha \beta }^{\gamma \delta }=(-)^{p(\beta
)(p(\alpha )+p(\gamma ))}\ A_{\alpha \gamma }B_{\beta \delta }  \label{gra.1}
\end{equation}
and the action of the permutation operator ${\cal P}$ on the vector $\left|
\alpha \right\rangle \stackrel{s}{\otimes }\left| \beta \right\rangle \in V%
\stackrel{s}{\otimes }V$ is given by 
\begin{equation}
{\cal P}\ \left| \alpha \right\rangle \stackrel{s}{\otimes }\left| \beta
\right\rangle =(-)^{p(\alpha )p(\beta )}\left| \beta \right\rangle \stackrel{%
s}{\otimes }\left| \alpha \right\rangle \Longrightarrow ({\cal P})_{\alpha
\beta }^{\gamma \delta }=(-)^{p(\alpha )p(\beta )}\delta _{\alpha \beta }\
\delta _{\gamma \delta },  \label{gra.2}
\end{equation}
where $p(\alpha )=1\ (0)$ if $\left| \alpha \right\rangle $ is an odd (even)
element.

Besides ${\cal R}$ we have to consider matrices $R={\cal PR}$ which satisfy 
\begin{equation}
R_{12}(u)R_{23}(u+v)R_{12}(v)=R_{23}(v)R_{12}(u+v)R_{23}(u).  \label{gra.2a}
\end{equation}

The regular solution of the graded {\small YB} equation for the fundamental
representation of $osp(2|1)$ algebra was found by Bazhanov and Shadrikov in 
\cite{BS}. It has the form 
\begin{equation}
R(u,\eta )=\left( 
\begin{array}{ccccccccccc}
x_{1} & 0 & 0 &  & 0 & 0 & 0 &  & 0 & 0 & 0 \\ 
0 & y_{5} & 0 &  & x_{2} & 0 & 0 &  & 0 & 0 & 0 \\ 
0 & 0 & y_{7} &  & 0 & y_{6} & 0 &  & x_{3} & 0 & 0 \\ 
&  &  &  &  &  &  &  &  &  &  \\ 
0 & x_{2} & 0 &  & x_{5} & 0 & 0 &  & 0 & 0 & 0 \\ 
0 & 0 & -y_{6} &  & 0 & -x_{4} & 0 &  & -x_{6} & 0 & 0 \\ 
0 & 0 & 0 &  & 0 & 0 & y_{5} &  & 0 & x_{2} & 0 \\ 
&  &  &  &  &  &  &  &  &  &  \\ 
0 & 0 & x_{3} &  & 0 & x_{6} & 0 &  & x_{7} & 0 & 0 \\ 
0 & 0 & 0 &  & 0 & 0 & x_{2} &  & 0 & x_{5} & 0 \\ 
0 & 0 & 0 &  & 0 & 0 & 0 &  & 0 & 0 & x_{1}
\end{array}
\right) ,  \label{gra.3}
\end{equation}
where 
\begin{eqnarray}
x_{1} &=&\sinh (u+2\eta )\sinh (u+3\eta ),\quad x_{2}=\sinh u\sinh (u+3\eta
),  \nonumber \\
x_{3} &=&\sinh u\sinh (u+\eta ),\quad x_{4}(u)=\sinh u\sinh (u+3\eta )-\sinh
2\eta \sinh 3\eta ,  \nonumber \\
x_{5} &=&{\rm e}^{-u}\sinh 2\eta \sinh (u+3\eta ),\qquad y_{5}={\rm e}%
^{u}\sinh 2\eta \sinh (u+3\eta ),  \nonumber \\
x_{6} &=&-\epsilon {\rm e}^{-u-2\eta }\sinh 2\eta \sinh u,\ \ \qquad
y_{6}=\epsilon {\rm e}^{u+2\eta }\sinh 2\eta \sinh u,  \nonumber \\
x_{7} &=&{\rm e}^{-u}\sinh 2\eta \ \left( \sinh (u+3\eta )+{\rm e}^{-\eta
}\sinh u\right) ,  \nonumber \\
y_{7} &=&{\rm e}^{u}\sinh 2\eta \ \left( \sinh (u+3\eta )+{\rm e}^{\eta
}\sinh u\right) .  \label{gra.4}
\end{eqnarray}
where $\epsilon =\pm 1$. Here we have assumed that the grading of threefold
space is $p(1)=p(3)=0$ and $p(2)=1$ and we will choose the solution with $%
\epsilon =1$.

Let us consider the inhomogeneous vertex model, where to each vertex we
associate two parameters: the global spectral parameter $u$ and the disorder
parameter $z$. In this case, the vertex weight matrix ${\cal R}$ depends on $%
u-z$ and consequently the monodromy matrix will be a function of the
disorder parameters $z_{i}$.

The graded quantum inverse scattering method is characterized by the
monodromy matrix $T(u|z)$ satisfying the equation 
\begin{equation}
R(u-v)\left[ T(u|z)\stackrel{s}{\otimes }T(v|z)\right] =\left[ T(v|z)%
\stackrel{s}{\otimes }T(u|z)\right] R(u-v),  \label{gra.5}
\end{equation}
whose consistency is guaranteed by the graded version of the {\small YB}
equation (\ref{int.1}). $T(u|z)$ is a matrix in the space $V$ (the auxiliary
space) whose matrix elements are operators on the states of the quantum
system (the quantum space, which will also be the space $V$). The monodromy
operator $T(u|z)$ is defined as an ordered product of local operators ${\cal %
L}_{n}$ (Lax operator), on all sites of the lattice: 
\begin{equation}
T(u|z)={\cal L}_{N}(u-z_{N}){\cal L}_{N-1}(u-z_{N-1})\cdots {\cal L}%
_{1}(u-z_{1}).  \label{gra.6}
\end{equation}
The Lax operator on the $n^{th}$ quantum space is given the normalized
graded permutation of (\ref{gra.3}): 
\begin{eqnarray}
{\cal L}_{n} &=&\frac{1}{x_{2}}\left( 
\begin{array}{ccccccccccc}
x_{1} & 0 & 0 &  & 0 & 0 & 0 &  & 0 & 0 & 0 \\ 
0 & x_{2} & 0 &  & x_{5} & 0 & 0 &  & 0 & 0 & 0 \\ 
0 & 0 & x_{3} &  & 0 & x_{6} & 0 &  & x_{7} & 0 & 0 \\ 
&  &  &  &  &  &  &  &  &  &  \\ 
0 & y_{5} & 0 &  & x_{2} & 0 & 0 &  & 0 & 0 & 0 \\ 
0 & 0 & y_{6} &  & 0 & x_{4} & 0 &  & x_{6} & 0 & 0 \\ 
0 & 0 & 0 &  & 0 & 0 & x_{2} &  & 0 & x_{5} & 0 \\ 
&  &  &  &  &  &  &  &  &  &  \\ 
0 & 0 & y_{7} &  & 0 & y_{6} & 0 &  & x_{3} & 0 & 0 \\ 
0 & 0 & 0 &  & 0 & 0 & y_{5} &  & 0 & x_{2} & 0 \\ 
0 & 0 & 0 &  & 0 & 0 & 0 &  & 0 & 0 & x_{1}
\end{array}
\right)  \nonumber \\
&=&\left( 
\begin{array}{lll}
L_{11}^{(n)}(u-z_{n}) & L_{12}^{(n)}(u-z_{n}) & L_{13}^{(n)}(u-z_{n}) \\ 
L_{21}^{(n)}(u-z_{n}) & L_{22}^{(n)}(u-z_{n}) & L_{23}^{(n)}(u-z_{n}) \\ 
L_{31}^{(n)}(u-z_{n}) & L_{32}^{(n)}(u-z_{n}) & L_{33}^{(n)}(u-z_{n})
\end{array}
\right)  \label{gra.7}
\end{eqnarray}
Note that $L_{\alpha \beta }^{(n)}(u),\ \alpha ,\beta =1,2,3$ are $3$ by $3$
matrices acting on the $n^{th}$ site of the lattice. It means that the
monodromy matrix has the form 
\begin{equation}
T(u|z)=\left( 
\begin{array}{lll}
A_{1}(u|z) & B_{1}(u|z) & B_{2}(u|z) \\ 
C_{1}(u|z) & A_{2}(u|z) & B_{3}(u|z) \\ 
C_{2}(u|z) & C_{3}(u|z) & A_{3}(u|z)
\end{array}
\right) ,  \label{gra.8}
\end{equation}
where 
\begin{eqnarray}
T_{ij}(u|z) &=&\sum_{k_{1},...,k_{N-1}=1}^{3}L_{ik_{1}}^{(N)}(u-z_{N})%
\stackrel{s}{\otimes }L_{k_{1}k_{2}}^{(N-1)}(u-z_{N-1})\stackrel{s}{\otimes }%
\cdots \stackrel{s}{\otimes }L_{k_{N-1}j}^{(1)}(u-z_{1}).  \nonumber \\
i,j &=&1,2,3.  \label{gra.9}
\end{eqnarray}

The vector in the quantum space of the monodromy matrix $T(u|z)$ that is
annihilated by the operators $T_{ij}(u|z)$, $i>j$ ($C_{i}(u|z)$ operators, $%
i=1,2,3$) and it is also an eigenvector for the operators $T_{ii}(u|z)$ ( $%
A_{i}(u|z)$ operators, $i=1,2,3$) is called a highest vector of the
monodromy matrix $T(u|z)$.

The transfer matrix $\tau (u|z)$ of the corresponding integrable spin model
is given by the supertrace of the monodromy matrix in the space $V$ 
\begin{equation}
\tau (u|z)=\sum_{i=1}^{3}(-1)^{p(i)}\
T_{ii}(u|z)=A_{1}(u|z)-A_{2}(u|z)+A_{3}(u|z).  \label{gra.10}
\end{equation}

Algebraic Bethe Ansatz solution for the inhomogeneous $osp(2|1)$ vertex
model can be obtained from the homogeneous case \cite{LI2}. The only
modification is a local shift of the spectral parameter $u\rightarrow
u-z_{i} $.

Here we will define some functions that will be used in the calculations of
the Bethe Ansatz:

\begin{eqnarray}
z(u) &=&\frac{x_{1}(u)}{x_{2}(u)}=\frac{\sinh (u+2\eta )}{\sinh u},\quad
y(u)=\frac{x_{3}(u)}{y_{6}(u)}=\frac{\sinh (u+\eta )}{{\rm e}^{u+2\eta
}\sinh 2\eta },\quad  \nonumber \\
&&  \nonumber \\
\omega (u) &=&-\frac{x_{1}(u)x_{3}(u)}{x_{4}(u)x_{3}(u)-x_{6}(u)y_{6}(u)}=-%
\frac{\sinh (u+\eta )}{\sinh (u-\eta )},  \nonumber \\
&&  \nonumber \\
{\cal Z}(u_{k}-u_{j}) &=&\left\{ 
\begin{array}{c}
z(u_{k}-u_{j})\qquad \qquad \quad \quad {\rm if}\quad k>j \\ 
z(u_{k}-u_{j})\omega (u_{j}-u_{k})\quad \ {\rm if}\quad k<j
\end{array}
\right. .  \label{inh.1}
\end{eqnarray}
We start defining the highest vector of the monodromy matrix $T(u|z)$ in a
lattice of $N$ sites as the even (bosonic) completely unoccupied state 
\begin{equation}
\left| 0\right\rangle =\otimes _{a=1}^{N}\left( 
\begin{array}{c}
1 \\ 
0 \\ 
0
\end{array}
\right) _{a}.  \label{inh.2}
\end{equation}
Using (\ref{gra.9}) we can compute the normalized action of the monodromy
matrix entries on this state 
\begin{eqnarray}
A_{i}(u|z)\left| 0\right\rangle &=&X_{i}(u|z)\left| 0\right\rangle ,\quad
C_{i}(u|z)\left| 0\right\rangle =0,\quad B_{i}(u|z)\left| 0\right\rangle
\neq \left\{ 0,\left| 0\right\rangle \right\} ,  \nonumber \\
X_{i}(u|z) &=&\prod_{a=1}^{N}\frac{x_{i}(u-z_{a})}{x_{2}(u-z_{a})},\qquad
i=1,2,3.  \label{inh.3}
\end{eqnarray}

The Bethe vectors are defined as normal ordered states $\Psi
_{n}(u_{1},\cdots ,u_{n})$ which can be written with aid of a recurrence
formula \cite{TA}: 
\begin{eqnarray}
&&\left. \Psi _{n}(u_{1},...,u_{n}|z)=B_{1}(u_{1}|z)\Psi
_{n-1}(u_{2},...,u_{n}|z)\right.  \nonumber \\
&&  \nonumber \\
&&\left. -B_{2}(u_{1}|z)\sum_{j=2}^{n}\frac{X_{1}(u_{j}|z)}{y(u_{1}-u_{j})}%
\prod_{k=2,k\neq j}^{n}{\cal Z}(u_{k}-u_{j})\Psi _{n-2}(u_{2},...,\stackrel{%
\wedge }{u}_{j},...,u_{n}|z)\right. ,  \label{inh.5}
\end{eqnarray}
with the initial condition $\Psi _{0}=\left| 0\right\rangle ,\quad \Psi
_{1}(u_{1}|z)=B_{1}(u_{1}|z)\left| 0\right\rangle $. Here \ $\stackrel{%
\wedge }{u}_{j}$ denotes that the rapidity $u_{j}$ is absent: $\Psi (%
\stackrel{\wedge }{u}_{j}|z)=\Psi (u_{1},...,u_{j-1},u_{j+1},\cdots
,u_{n}|z) $.

The action of the transfer matrix $\tau (u|z)$ on these Bethe vectors gives
us the following off-shell Bethe Ansatz equation for the $osp(2|1)$ vertex
model

\begin{equation}
\tau (u|z)\Psi _{n}(u_{1},...,u_{n}|z)=\Lambda _{n}\Psi
_{n}(u_{1},...,u_{n}|z)-\sum_{j=1}^{n}{\cal F}_{j}^{(n-1)}\Psi
_{(n-1)}^{j}+\sum_{j=2}^{n}\sum_{l=1}^{j-1}{\cal F}_{lj}^{(n-2)}\Psi
_{(n-2)}^{lj}.  \label{inh.4}
\end{equation}

Let us now describe each term which appear in the right hand side of (\ref
{inh.4}) ( for more details the reader can see \cite{LI2}): In the first
term the Bethe vectors (\ref{inh.5}) are multiplied by $c$-numbers $\Lambda
_{n}=\Lambda _{n}(u,u_{1},...,u_{n}|z)$ given by 
\begin{equation}
\Lambda
_{n}=X_{1}(u|z)\prod_{k=1}^{n}z(u_{k}-u)-(-)^{n}X_{2}(u|z)\prod_{k=1}^{n}%
\frac{z(u-u_{k})}{\omega (u-u_{k})}+X_{3}(u|z)\prod_{k=1}^{n}\frac{%
x_{2}(u-u_{k})}{x_{3}(u-u_{k})}.  \label{inh.6}
\end{equation}
The second term is a sum of new vectors 
\begin{equation}
\Psi _{(n-1)}^{j}=\left( \frac{x_{5}(u_{j}-u)}{x_{2}(u_{j}-u)}B_{1}(u|z)+%
\frac{1}{y(u-u_{j})}B_{3}(u|z)\right) \Psi _{n-1}(\stackrel{\wedge }{u}_{j}),
\label{inh.7}
\end{equation}
multiplied by scalar functions ${\cal F}_{j}^{(n-1)}$ given by 
\begin{equation}
{\cal F}_{j}^{(n-1)}=X_{1}(u_{j}|z)\prod_{k\neq j}^{n}{\cal Z}%
(u_{k}-u_{j})+(-)^{n}X_{2}(u_{j}|z)\prod_{k\neq j}^{n}{\cal Z}(u_{j}-u_{k}).
\label{inh.8}
\end{equation}
Finally, the last term is a coupled sum of a third type of vector-valued
functions 
\begin{equation}
\Psi _{(n-2)}^{lj}=B_{2}(u|z)\Psi _{n-2}(\stackrel{\wedge }{u}_{l},\stackrel{%
\wedge }{u}_{j}),  \label{inh.9}
\end{equation}
with intricate coefficients 
\begin{eqnarray}
{\cal F}_{lj}^{(n-2)} &=&G_{lj}X_{1}(u_{l}|z)X_{1}(u_{j}|z)\prod_{k=1,k\neq
j,l}^{n}{\cal Z}(u_{k}-u_{l}){\cal Z}(u_{k}-u_{j})  \nonumber \\
&&-(-)^{n}Y_{lj}X_{1}(u_{l}|z)X_{2}(u_{j}|z)\prod_{k=1,k\neq j,l}^{n}{\cal Z}%
(u_{k}-u_{l}){\cal Z}(u_{j}-u_{k})  \nonumber \\
&&-(-)^{n}F_{lj}X_{1}(u_{j}|z)X_{2}(u_{l}|z)\prod_{k=1,k\neq j,l}^{n}{\cal Z}%
(u_{l}-u_{k}){\cal Z}(u_{k}-u_{j})  \nonumber \\
&&+H_{lj}X_{2}(u_{l}|z)X_{2}(u_{j}|z)\prod_{k=1,k\neq j,l}^{n}{\cal Z}%
(u_{j}-u_{k}){\cal Z}(u_{l}-u_{k}).  \label{inh.10}
\end{eqnarray}
where $G_{lj}$ , $Y_{lj}$ , $F_{lj}$ and $H_{lj}$ are additional ratio
functions defined by 
\begin{eqnarray}
G_{lj} &=&\frac{x_{7}(u_{l}-u)}{x_{3}(u_{l}-u)}\frac{1}{y(u_{l}-u_{j})}+%
\frac{z(u_{l}-u)}{\omega (u_{l}-u)}\frac{x_{5}(u_{j}-u)}{x_{2}(u_{j}-u)}%
\frac{1}{y(u-u_{l})},  \nonumber \\
&&  \nonumber \\
H_{lj} &=&\frac{y_{7}(u-u_{l})}{x_{3}(u-u_{l})}\frac{1}{y(u_{l}-u_{j})}-%
\frac{y_{5}(u-u_{l})}{x_{3}(u-u_{l})}\frac{1}{y(u-u_{j})},  \nonumber \\
&&  \nonumber \\
Y_{lj} &=&\frac{1}{y(u-u_{l})}\left\{ z(u-u_{l})\frac{y_{5}(u-u_{j})}{%
x_{2}(u-u_{j})}-\frac{y_{5}(u-u_{l})}{x_{2}(u-u_{l})}\frac{y_{5}(u_{l}-u_{j})%
}{x_{2}(u_{l}-u_{j})}\right\} ,  \nonumber \\
&&  \nonumber \\
F_{lj} &=&\frac{y_{5}(u-u_{l})}{x_{2}(u-u_{l})}\left\{ \frac{%
y_{5}(u_{l}-u_{j})}{x_{2}(u_{l}-u_{j})}\frac{1}{y(u-u_{l})}+\frac{z(u-u_{l})%
}{\omega (u-u_{l})}\frac{1}{y(u-u_{j})}\right.  \nonumber \\
&&\left. -\frac{y_{5}(u-u_{l})}{x_{2}(u-u_{l})}\frac{1}{y(u_{l}-u_{j})}%
\right\} .  \label{inh.11}
\end{eqnarray}
In the usual Bethe Ansatz method, the next step consist in impose the
vanishing of the so-called unwanted terms of (\ref{inh.4}) in order to get
an eigenvalue problem for the transfer matrix:

We impose ${\cal F}_{j}^{(n-1)}=0$ and ${\cal F}_{lj}^{(n-2)}=0$ into (\ref
{inh.4}) to recover the eigenvalue problem. This means that $\Psi
_{n}(u_{1},...,u_{n}|z)$ is an eigenstate of $\tau (u|z)$ with eigenvalue $%
\Lambda _{n}$ , provided the rapidities $u_{j}$ are solutions of the
inhomogeneous Bethe Ansatz equations 
\begin{eqnarray}
\prod_{a=1}^{N}z(u_{j}-z_{a}) &=&(-)^{n+1}\prod_{k=1,\ k\neq j}^{n}\frac{%
z(u_{j}-u_{k})}{z(u_{k}-u_{j})}\omega (u_{k}-u_{j}),  \nonumber \\
j &=&1,2,...,n.  \label{inh.12}
\end{eqnarray}

\section{Structure of the ${\bf osp(2|1)}$ Gaudin Model}

In this section we will consider the theory of the Gaudin model. To do this
we need to calculate the semi-classical limit of the results presented in
the previous section.

In order to expand the matrix elements of $T(u|z)$, up to an appropriate
order in $\eta $, we will start by expanding the Lax operator entries
defined in (\ref{gra.7}): 
\begin{eqnarray}
L_{11}^{(n)} &=&{\cal I}_{n}+2\eta \coth (u-z_{n})\ {\cal H}_{n}+2\eta
^{2}\left( {\cal H}_{n}^{2}+\frac{3}{2}\frac{{\cal H}_{n}^{2}-{\cal H}_{n}}{%
\sinh (u-z_{n})^{2}}\right) +{\rm o}(\eta ^{3}),  \nonumber \\
&&  \nonumber \\
L_{22}^{(n)} &=&{\cal I}_{n}-2\eta ^{2}\frac{3\left( {\cal I}_{n}-{\cal H}%
_{n}^{2}\right) }{\sinh (u-z_{n})^{2}}+{\rm o}(\eta ^{3}),  \nonumber \\
&&  \nonumber \\
L_{33}^{(n)} &=&{\cal I}_{n}-2\eta \coth u\ {\cal H}_{n}+2\eta ^{2}\left( 
{\cal H}_{n}^{2}+\frac{3}{2}\frac{{\cal H}_{n}^{2}+{\cal H}_{n}}{\sinh
(u-z_{n})^{2}}\right) +{\rm o}(\eta ^{3}).  \label{gau.1}
\end{eqnarray}
and for the elements out of the diagonal we have 
\begin{eqnarray}
L_{12}^{(n)} &=&-2\eta \ \frac{{\rm e}^{-u+z_{n}}}{\sinh (u-z_{n})}{\cal V}%
_{n}^{-}+{\rm o}(\eta ^{2}),\quad \ L_{21}^{(n)}=2\eta \ \frac{{\rm e}%
^{u-z_{n}}}{\sinh (u-z_{n})}{\cal V}_{n}^{+}+{\rm o}(\eta ^{2}),  \nonumber
\\
&&  \nonumber \\
L_{23}^{(n)} &=&2\eta \ \frac{{\rm e}^{-u+z_{n}}}{\sinh (u-z_{n})}{\cal V}%
_{n}^{-}+{\rm o}(\eta ^{2}),\quad \quad L_{32}^{(n)}=2\eta \ \frac{{\rm e}%
^{u-z_{n}}}{\sinh (u-z_{n})}{\cal V}_{n}^{+}+{\rm o}(\eta ^{2}),  \nonumber
\\
&&  \nonumber \\
L_{13}^{(n)} &=&2\eta \ \frac{{\rm e}^{-u+z_{n}}}{\sinh (u-z_{n})}{\cal X}%
_{n}^{-}+{\rm o}(\eta ^{2}),\quad \quad L_{31}^{(n)}=2\eta \ \frac{{\rm e}%
^{u-z_{n}}}{\sinh (u-z_{n})}{\cal X}_{n}^{+}+{\rm o}(\eta ^{2}).
\label{gau.2}
\end{eqnarray}
where ${\cal V}^{\pm }=2V^{\pm }$, ${\cal X}^{\pm }=2X^{\pm }$ and ${\cal H}%
=2H$.

Substituting (\ref{gau.1}) and (\ref{gau.2}) into (\ref{gra.9}) we will get
the semi-classical expansion for the monodromy matrix entries: 
\begin{eqnarray}
A_{1}(u|z) &=&{\cal I}+2\eta \sum_{a=1}^{N}\coth (u-z_{a}){\cal H}_{a}+4\eta
^{2}\left\{ \sum_{a<b}\coth (u-z_{a})\coth (u-z_{b}){\cal H}_{a}\stackrel{s}{%
\otimes }{\cal H}_{b}\right.  \nonumber \\
&&+\left. \sum_{a<b}\frac{{\rm e}^{z_{a}-z_{b}}}{\sinh (u-z_{a})\sinh
(u-z_{b})}\left( {\cal X}_{a}^{-}\stackrel{s}{\otimes }{\cal X}_{b}^{+}-%
{\cal V}_{a}^{-}\stackrel{s}{\otimes }{\cal V}_{b}^{+}\right) \right. 
\nonumber \\
&&+\frac{1}{2}\left. \sum_{a=1}^{N}\left( {\cal H}_{a}^{2}+\frac{3}{2}\frac{%
{\cal H}_{a}^{2}-{\cal H}_{a}}{\sinh (u-z_{a})^{2}}\right) \right\} +{\rm o}%
(\eta ^{3}),  \label{gau.3a}
\end{eqnarray}
\begin{eqnarray}
A_{2}(u|z) &=&{\cal I}-4\eta ^{2}\left\{ \sum_{a<b}\frac{{\rm e}%
^{-z_{a}+z_{b}}{\cal V}_{a}^{+}\stackrel{s}{\otimes }{\cal V}_{b}^{-}-{\rm e}%
^{z_{a}-z_{b}}{\cal V}_{a}^{-}\stackrel{s}{\otimes }{\cal V}_{b}^{+}}{\sinh
(u-z_{a})\sinh (u-z_{b})}+\frac{3}{2}\sum_{a=1}^{N}\frac{{\cal I}_{a}-{\cal H%
}_{a}^{2}}{\sinh (u-z_{a})^{2}}\right\}  \nonumber \\
&&+{\rm o}(\eta ^{3}),  \label{gau.3b}
\end{eqnarray}
\begin{eqnarray}
A_{3}(u|z) &=&{\cal I}-2\eta \sum_{a=1}^{N}\coth (u-z_{a}){\cal H}_{a}+4\eta
^{2}\left\{ \sum_{a<b}\coth (u-z_{a})\coth (u-z_{b}){\cal H}_{a}\stackrel{s}{%
\otimes }{\cal H}_{b}\right.  \nonumber \\
&&+\left. \sum_{a<b}\frac{{\rm e}^{-z_{a}+z_{b}}}{\sinh (u-z_{a})\sinh
(u-z_{b})}\left( {\cal X}_{a}^{+}\stackrel{s}{\otimes }{\cal X}_{b}^{-}+%
{\cal V}_{a}^{+}\stackrel{s}{\otimes }{\cal V}_{b}^{-}\right) \right. 
\nonumber \\
&&+\frac{1}{2}\left. \sum_{a=1}^{N}\left( {\cal H}_{a}^{2}+\frac{3}{2}\frac{%
{\cal H}_{a}^{2}+{\cal H}_{a}}{\sinh (u-z_{a})^{2}}\right) \right\} +{\rm o}%
(\eta ^{3}).  \label{gau.3c}
\end{eqnarray}
For off-diagonal elements we only need to expand them up to the first order
in $\eta $%
\begin{eqnarray}
B_{1}(u|z) &=&-B_{3}(u|z)=-2\eta \sum_{a=1}^{N}\frac{{\rm e}^{-u+z_{a}}}{%
\sinh (u-z_{a})}{\cal V}_{a}^{-}+{\rm o}(\eta ^{2}),  \nonumber \\
C_{1}(u|z) &=&C_{3}(u|z)=2\eta \sum_{a=1}^{N}\frac{{\rm e}^{u-z_{a}}}{\sinh
(u-z_{a})}{\cal V}_{a}^{+}+{\rm o}(\eta ^{2}),  \nonumber \\
B_{2}(u|z) &=&2\eta \sum_{a=1}^{N}\frac{{\rm e}^{-u+z_{a}}}{\sinh (u-z_{a})}%
{\cal X}_{a}^{-}+{\rm o}(\eta ^{2}),  \nonumber \\
C_{2}(u|z) &=&2\eta \sum_{a=1}^{N}\frac{{\rm e}^{u-z_{a}}}{\sinh (u-z_{a})}%
{\cal X}_{a}^{+}+{\rm o}(\eta ^{2}).  \label{gau.4}
\end{eqnarray}
Therefore, the semi-classical expansion of the transfer matrix (\ref{gra.10}%
) has the form 
\begin{equation}
\tau (u|z)={\cal I}+8\eta ^{2}\tau ^{(2)}(u|z)+{\rm o}(\eta ^{3}),
\label{gau.5}
\end{equation}
where 
\begin{equation}
\tau ^{(2)}(u|z)=\sum_{a=1}^{N}\frac{{\cal G}_{a}(u)}{{\rm e}^{u-z_{a}}\sinh
(u-z_{a})}+\sum_{a=1}^{N}\left( \frac{1}{2}{\cal H}_{a}^{2}+\frac{3}{4}\frac{%
{\cal I}_{a}}{\sinh (u-z_{a})^{2}}\right) ,  \label{gau.5a}
\end{equation}
with 
\begin{eqnarray}
{\cal G}_{a}(u) &=&\sum_{b\neq a}\frac{1}{\sinh (z_{a}-z_{b})}\left\{ \cosh
(u-z_{a})\cosh (u-z_{b}){\cal H}_{a}\stackrel{s}{\otimes }{\cal H}_{b}\right.
\nonumber \\
&&+\left. \frac{1}{2}\left( {\rm e}^{-z_{a}+z_{b}}{\cal X}_{a}^{+}\stackrel{s%
}{\otimes }{\cal X}_{b}^{-}+{\rm e}^{z_{a}-z_{b}}{\cal X}_{a}^{-}\stackrel{s%
}{\otimes }{\cal X}_{b}^{+}\right) \right.  \nonumber \\
&&+\left. \left( {\rm e}^{-z_{a}+z_{b}}{\cal V}_{a}^{+}\stackrel{s}{\otimes }%
{\cal V}_{b}^{-}-{\rm e}^{z_{a}-z_{b}}{\cal V}_{a}^{-}\stackrel{s}{\otimes }%
{\cal V}_{b}^{+}\right) \right\} .  \label{gau.6}
\end{eqnarray}
Here we have used the symmetry 
\begin{equation}
{\cal G}_{ba}(u,z_{a},z_{b})={\cal P\ }{\cal G}_{ab}(u,z_{a},z_{b})\ {\cal P}%
={\cal G}_{ab}(u,z_{b},z_{a}),  \label{gau.7}
\end{equation}
and the identity 
\begin{equation}
\frac{1}{\sinh (u-z_{a})\sinh (u-z_{b})}=\frac{1}{\sinh (z_{a}-z_{b})}\left( 
\frac{{\rm e}^{-u+z_{a}}}{\sinh (u-z_{a})}-\frac{{\rm e}^{-u+z_{b}}}{\sinh
(u-z_{b})}\right) .  \label{gau.8}
\end{equation}

The Gaudin Hamiltonians are defined as the residue of $\tau (u|z)$ at the
point $u=z_{a}$. This results in $N$ non-local Hamiltonians 
\begin{eqnarray}
G_{a} &=&\sum_{b\neq a}^{N}\frac{1}{\sinh (z_{a}-z_{b})}\left\{ \cosh
(z_{a}-z_{b}){\cal H}_{a}\stackrel{s}{\otimes }{\cal H}_{b}+\frac{1}{2}%
\left( {\rm e}^{-z_{a}+z_{b}}{\cal X}_{a}^{+}\stackrel{s}{\otimes }{\cal X}%
_{b}^{-}\right. \right.  \nonumber \\
&&+\left. {\rm e}^{z_{a}-z_{b}}{\cal X}_{a}^{-}\stackrel{s}{\otimes }{\cal X}%
_{b}^{+}\right) +\left. {\rm e}^{-z_{a}+z_{b}}{\cal V}_{a}^{+}\stackrel{s}{%
\otimes }{\cal V}_{b}^{-}-{\rm e}^{z_{a}-z_{b}}{\cal V}_{a}^{-}\stackrel{s}{%
\otimes }{\cal V}_{b}^{+}\right\} ,  \nonumber \\
a &=&1,2,...,N.  \label{gau.9}
\end{eqnarray}
satisfying 
\begin{equation}
\sum_{a=1}^{N}G_{a}=0,\quad \frac{\partial G_{a}}{\partial z_{b}}=\frac{%
\partial G_{b}}{\partial z_{a}},\quad \left[ G_{a},G_{b}\right] =0,\qquad
\forall a,b.  \label{gau.10}
\end{equation}

In the next section we will use the data of the algebraic Bethe Ansatz for
the $osp(2|1)$ vertex model to find the exact spectrum and eigenvectors for
each of these $N-1$ independent Hamiltonians.

Before doing this, we would like to consider the semi-classical limit of the
fundamental commutation relation (\ref{gra.5}) in order to get the $osp(2|1)$
Gaudin algebra:

The semi-classical expansions of $T$ and $R$ can be written in the following
form 
\begin{equation}
T(u|z)=1+2\eta l(u|z)+o(\eta ^{2}),\quad R(u)={\cal P}\left[ 1+2\eta
r(u)+o(\eta ^{2})\right] .  \label{gau.11}
\end{equation}
From (\ref{gau.3a}--\ref{gau.4}) we can see that the \ ''classical $l$%
-operator '' has the form 
\begin{equation}
l(u|z)=\left( 
\begin{array}{ccc}
{\cal H}(u|z) & -{\cal V}^{-}(u|z) & {\cal X}^{-}(u|z) \\ 
{\cal V}^{+}(u|z) & 0 & {\cal V}^{-}(u|z) \\ 
{\cal X}^{+}(u|z) & {\cal V}^{+}(u|z) & -{\cal H}(u|z)
\end{array}
\right) ,  \label{gau.12}
\end{equation}
where 
\begin{eqnarray}
{\cal H}(u|z) &=&\sum_{a=1}^{N}\coth (u-z_{a}){\cal H}_{a},  \nonumber \\
{\cal V}^{-}(u|z) &=&\sum_{a=1}^{N}\frac{{\rm e}^{-u+z_{a}}}{\sinh (u-z_{a})}%
{\cal V}_{a}^{-},\qquad {\cal V}^{+}(u|z)=\sum_{a=1}^{N}\frac{{\rm e}%
^{u-z_{a}}}{\sinh (u-z_{a})}{\cal V}_{a}^{+},  \nonumber \\
{\cal X}^{-}(u|z) &=&\sum_{a=1}^{N}\frac{{\rm e}^{-u+z_{a}}}{\sinh (u-z_{a})}%
{\cal X}_{a}^{-},\qquad {\cal X}^{+}(u|z)=\sum_{a=1}^{N}\frac{{\rm e}%
^{u-z_{a}}}{\sinh (u-z_{a})}{\cal X}_{a}^{+}.  \label{gau.13}
\end{eqnarray}
The classical $r$-matrix has the form 
\begin{eqnarray}
r(u) &=&\frac{1}{\sinh u}\left\{ \cosh u\ {\cal H}\stackrel{s}{\otimes }%
{\cal H}+\frac{1}{2}\left( {\rm e}^{-u}\ {\cal X}^{+}\stackrel{s}{\otimes }%
{\cal X}^{-}+{\rm e}^{u}\ {\cal X}^{-}\stackrel{s}{\otimes }{\cal X}%
^{+}\right) \right.  \nonumber \\
&&+\left. {\rm e}^{-u}\ ({\cal HV}^{+}+{\cal V}^{+}{\cal H})\stackrel{s}{%
\otimes }({\cal V}^{-}{\cal H}+{\cal HV}^{-})\right.  \nonumber \\
&&-\left. {\rm e}^{u}\ ({\cal V}^{-}{\cal H}+{\cal HV}^{-})\stackrel{s}{%
\otimes }({\cal HV}^{+}+{\cal V}^{+}{\cal H})\right\} .  \label{gau.14}
\end{eqnarray}
Here we observe that we are in the fundamental representation of the $%
osp(2|1)$ algebra where the relation 
\begin{equation}
({\cal HV}^{+}+{\cal V}^{+}{\cal H})({\cal V}^{-}{\cal H}+{\cal HV}^{-})=%
{\cal V}^{+}{\cal V}^{-}  \label{gau.15}
\end{equation}
holds. Therefore,\ (\ref{gau.14}) is equivalent to that $r$-matrix
constructed out of the quadratic Casimir in a standard way \cite{Ku2}.
Indeed it corresponds to the second regular solution ($\epsilon =-1)$
presented in (\ref{gra.4}).

Substituting (\ref{gau.14}) and (\ref{gau.12}) into the fundamental relation
(\ref{gra.5}), we have 
\begin{eqnarray}
&&{\cal P}l(u|z)\stackrel{s}{\otimes }l(v|z)+{\cal P}r(u-v)\left[ l(u|z)%
\stackrel{s}{\otimes }1+1\stackrel{s}{\otimes }l(v|z)\right]  \nonumber \\
&=&l(v|z)\stackrel{s}{\otimes }l(u|z){\cal P}+\left[ l(v|z)\stackrel{s}{%
\otimes }1+1\stackrel{s}{\otimes }l(u|z)\right] {\cal P}r(u-v),
\label{gau.16}
\end{eqnarray}
whose consistence is guaranteed by the graded classical {\small YB} equation.

From (\ref{gau.16}) we can derive (anti)commutation relations between the
matrix elements of $l(u|z)$. They are the defining relations of the $%
osp(2|1) $ Gaudin algebra : 
\begin{eqnarray}
\lbrack {\cal H}(u|z),{\cal H}(v|z)] &=&0,  \nonumber \\
\lbrack {\cal V}^{\mp }(u|z),{\cal X}^{\mp }(v|z)] &=&[{\cal X}^{\mp }(u|z),%
{\cal X}^{\mp }(v|z)]=0,  \nonumber \\
\lbrack {\cal V}^{\pm }(u|z),{\cal X}^{\mp }(v|z)] &=&\frac{2{\rm e}^{\pm
(u-v)}}{\sinh (u-v)}\left[ {\cal V}^{\mp }(u|z)-{\cal V}^{\mp }(v|z)\right] ,
\nonumber \\
\lbrack {\cal X}^{-}(u|z),{\cal X}^{+}(v|z)] &=&\frac{4{\rm e}^{-u+v}}{\sinh
(u-v)}\left[ {\cal H}(u|z)-{\cal H}(v|z)\right] ,  \nonumber \\
\{{\cal V}^{-}(u|z),{\cal V}^{+}(v|z)\} &=&\frac{{\rm e}^{-u+v}}{\sinh (u-v)}%
\left[ {\cal H}(u|z)-{\cal H}(v|z))\right] ,  \nonumber \\
\lbrack {\cal H}(u|z),{\cal V}^{\mp }(v|z)] &=&\pm \frac{1}{\sinh (u-v)}%
\left[ {\rm e}^{\pm (u-v)}{\cal V}^{\mp }(u|z)-\cosh (u-v){\cal V}^{\mp
}(v|z)\right] ,  \nonumber \\
\lbrack {\cal H}(u|z),{\cal X}^{\mp }(v|z)] &=&\pm \frac{2}{\sinh (u-v)}%
\left[ {\rm e}^{\pm (u-v)}{\cal X}^{\mp }(u|z)-\cosh (u-v){\cal X}^{\mp
}(v|z)\right] ,  \nonumber \\
\{{\cal V}^{\mp }(u|z),{\cal V}^{\mp }(v|z)\} &=&\pm \frac{1}{\sinh (u-v)}%
\left[ {\rm e}^{\pm (u-v)}{\cal X}^{\mp }(u|z)-{\rm e}^{\mp (u-v)}{\cal X}%
^{\mp }(v|z))\right] .  \label{gau.17}
\end{eqnarray}
A direct consequence of these relations is the commutativity of $\tau
^{(2)}(u|z)$ 
\begin{equation}
\lbrack \tau ^{(2)}(u|z),\tau ^{(2)}(v|z)]=0,\qquad \forall u,v
\label{gau.18}
\end{equation}
from which the commutativity of the Gaudin Hamiltonians $G_{a}$ follows
immediately.

\subsection{Off-shell Gaudin Equation}

In order to get semi-classical limit of the {\small OSBAE} (\ref{inh.4}) we
first consider the semi-classical expansions of the Bethe vectors defined in
(\ref{inh.5}), (\ref{inh.7}) and (\ref{inh.9}): 
\begin{eqnarray}
\Psi _{n}(u_{1},...,u_{n}|z) &=&(-2\eta )^{n}\Phi _{n}(u_{1},...,u_{n}|z)+%
{\rm o}(\eta ^{n+1}),  \nonumber \\
&&  \nonumber \\
\Psi _{(n-1)}^{j} &=&2(-2\eta )^{n+1}\frac{{\rm e}^{u-u_{j}}}{\sinh (u-u_{j})%
}{\cal V}^{-}(u|z)\Phi _{n-1}(\stackrel{\wedge }{u}_{j}|z)+{\rm o}(\eta
^{n+2}),  \nonumber \\
&&  \nonumber \\
\Psi _{(n-2)}^{lj} &=&-(-2\eta )^{n-1}{\cal X}^{-}(u|z)\Phi _{n-2}(\stackrel{%
\wedge }{u}_{l},\stackrel{\wedge }{u}_{j}|z)+{\rm o}(\eta ^{n}),
\label{off.1}
\end{eqnarray}
where 
\begin{eqnarray}
\Phi _{n}(u_{1},...,u_{n}|z) &=&{\cal V}^{-}(u_{1}|z)\Phi
_{n-1}(u_{2},...,u_{n}|z)  \nonumber \\
&&-{\cal X}^{-}(u_{1}|z)\sum_{j=2}^{n}\frac{(-)^{j}{\rm e}^{u_{1}-u_{j}}}{%
\sinh (u_{1}-u_{j})}\Phi _{n-2}(\stackrel{\wedge }{u}_{j}|z),  \label{off.2}
\end{eqnarray}
with $\Phi _{0}=\left| 0\right\rangle $ and $\Phi _{1}(u_{1}|z)={\cal V}%
^{-}(u_{1}|z)\Phi _{0}$.

Here we would like make a few comments on the structure of these
vector-valued functions. In (\ref{off.2}) they are written in a normal
ordered form. Since we are working with fermionic degree of freedom, the
function $\Phi _{n}(u_{1},...,u_{n}|z)$ is totally antisymmetric. 
\begin{equation}
\Phi _{n}(u_{1},...,u_{i-1},u_{i+1},u_{i},...,u_{n}|z)=-\Phi
_{n}(u_{1},...,u_{i-1},u_{i},u_{i+1},...,u_{n}|z).  \label{off.3}
\end{equation}
To see this one can use the Gaudin algebra (\ref{gau.17}). For instance, in
its antisymmetric form the Bethe vector $\Phi _{2}$ reads as 
\begin{eqnarray}
\Phi _{2}(u_{1},u_{2}|z) &=&\frac{1}{2}[{\cal V}^{-}(u_{1}|z){\cal V}%
^{-}(u_{2}|z)-{\cal V}^{-}(u_{2}|z){\cal V}^{-}(u_{1}|z)]\Phi _{0}  \nonumber
\\
&&-\frac{1}{2}[\frac{{\rm e}^{u_{1}-u_{2}}{\cal X}^{-}(u_{1}|z)+{\rm e}%
^{-u_{1}+u_{2}}{\cal X}^{-}(u_{2}|z)}{\sinh (u_{1}-u_{2})}]\Phi _{0}.
\label{off.4}
\end{eqnarray}

Now we will consider the semi-classical expansions of the $c$-number
functions presented in the {\small OSBAE} (\ref{inh.4}) 
\begin{eqnarray}
\Lambda _{n} &=&1+2(-2\eta )^{2}\Lambda _{n}^{(2)}+{\rm o}(\eta ^{3}),\qquad 
{\cal F}_{j}^{(n-1)}=(-2\eta )\ (-)^{j+1}\ f_{j}^{(n-1)}+{\rm o}(\eta ^{2}) 
\nonumber \\
&&  \nonumber \\
{\cal F}_{lj}^{(n-2)} &=&2(-2\eta )^{3}\frac{(-)^{l+j+1}}{\sinh (u_{j}-u_{l})%
}\left\{ \frac{{\rm e}^{u-u_{j}}}{\sinh (u-u_{l})}f_{l}^{(n-1)}+\frac{{\rm e}%
^{u-u_{l}}}{\sinh (u-u_{j})}f_{j}^{(n-1)}\right\} +{\rm o}(\eta ^{4}), 
\nonumber \\
&&  \label{off.5}
\end{eqnarray}
\ where 
\begin{eqnarray}
\Lambda _{n}^{(2)} &=&\frac{1}{2}(N+n)+\frac{3}{4}\sum_{a=1}^{N}\frac{1}{%
\sinh (u-z_{a})^{2}}-\sum_{a=1}^{N}\sum_{j=1}^{n}\coth (u-z_{a})\coth
(u-u_{j})  \nonumber \\
&&+\sum_{a<b}^{N}\coth (u-z_{a})\coth (u-z_{b})+\sum_{j<k}^{n}\coth
(u-u_{j})\coth (u-u_{k}),  \label{off.6}
\end{eqnarray}
and 
\begin{equation}
f_{j}^{(n-1)}=-\sum_{a=1}^{N}\coth (u_{j}-z_{a})+\sum_{k\neq j}^{n}\coth
(u_{j}-u_{k}).  \label{off.7}
\end{equation}
Substituting these expressions into the (\ref{inh.4}) and comparing the
coefficients of the terms $2(-2\eta )^{n+2}$ we get the first non-trivial
consequence for the semi-classical limit of the \ {\small OSBAE}: 
\begin{equation}
\tau ^{(2)}(u|z)\ \Phi _{n}(u_{1},...,u_{n}|z)=\Lambda _{n}^{(2)}\ \Phi
_{n}(u_{1},...,u_{n}|z)-\sum_{j=1}^{n}\ \frac{(-)^{j}f_{j}^{(n-1)}\Theta
_{(n-1)}^{j}}{{\rm e}^{u_{j}-u}\sinh (u_{j}-u)}.  \label{off.8}
\end{equation}
Note that in this limit the contributions from $\Psi _{(n-1)}^{j}$ and $\Psi
_{(n-2)}^{lj}$ are combined to give a new vector valued function 
\begin{equation}
\Theta _{(n-1)}^{j}={\cal V}^{-}(u|z)\ \Phi _{n-1}(\ \stackrel{\wedge }{u}%
_{j}|z)-{\cal X}^{-}(u|z)\sum_{k=1,\ k\neq j}^{n}\frac{(-)^{k^{^{\prime }}}%
{\rm e}^{u_{j}-u_{k}}}{\sinh (u_{j}-u_{k})}\ \Phi _{n-2}(\ \stackrel{\wedge 
}{u}_{j},\stackrel{\wedge }{u}_{k}|z),  \label{off.9}
\end{equation}
where $k^{^{\prime }}=k+1\ \ $for$\quad k<j$ \ and $k^{^{\prime }}=k\ $\ for$%
\quad k>j$. Therefore, our graded {\small OSBAE} (\ref{off.8}) is very
similar to that presented by Babujian and Flume for simple Lie algebras \cite
{BF2}. This result could be expected since the superalgebra $osp(2|1)$ has
many features which make \ it very close to the Lie algebra \cite{FSS}.

Finally, we take the residue of (\ref{off.8}) at the point $u=z_{a}$ to get
the off-shell Gaudin equation: 
\begin{eqnarray}
G_{a}\Phi _{n}(u_{1},...,u_{n}|z) &=&g_{a}\Phi
_{n}(u_{1},...,u_{n}|z)-\sum_{l=1}^{n}\frac{(-)^{l}f_{l}^{(n-1)}\phi
_{(n-1)}^{l}}{{\rm e}^{u_{l}-z_{a}}\sinh (u_{l}-z_{a})},  \nonumber \\
a &=&1,2,...,N  \label{off.11}
\end{eqnarray}
where $g_{a}$ is the residue of $\Lambda _{n}^{(2)}$ 
\begin{equation}
g_{a}={\rm res}_{u=z_{a}}\Lambda _{n}^{(2)}=\sum_{b\neq a}^{N}\coth
(z_{a}-z_{b})-\sum_{l=1}^{n}\coth (z_{a}-u_{l}),  \label{off.12}
\end{equation}
and $\phi _{(n-1)}^{l}$ is the residue of $\Theta _{(n-1)}^{l}$ 
\begin{equation}
\phi _{(n-1)}^{j}={\rm res}_{u=z_{a}}\Theta _{(n-1)}^{j}={\cal V}%
_{a}^{-}\Phi _{n-1}(\stackrel{\wedge }{u}_{j}|z)-{\cal X}_{a}^{-}\sum_{k\neq
j}^{n}\frac{(-)^{k^{\prime }}{\rm e}^{u_{j}-u_{k}}}{\sinh (u_{j}-u_{k})}\Phi
_{n-2}(\stackrel{\wedge }{u}_{k},\stackrel{\wedge }{u}_{j}|z).
\label{off.13}
\end{equation}

In this way we are arriving to the main result of this paper. The equation (%
\ref{off.11}) permits us solve one of the main problem of the Gaudin model, 
{\it i.e.}, the determination of the eigenvalues and eigenvectors of the
commuting Hamiltonians $G_{a}$ (\ref{gau.6}): $g_{a}$ is the eigenvalue of $%
G_{a}$ with eigenfunction $\Phi _{n}$ provided $u_{l}$ are solutions of the
following equations $f_{j}^{(n-1)}=0$, {\it i.e}.: 
\begin{equation}
\sum_{k\neq j}^{n}\coth (u_{j}-u_{k})=\sum_{a=1}^{N}\coth
(u_{j}-z_{a}),\quad j=1,2,...,n.  \label{off.14}
\end{equation}
Moreover, as we will see in the next section, the off-shell Gaudin equation (%
\ref{off.11}) provides solutions for the differential equations known as 
{\small KZ} equations.

\section{Knizhnik-Zamolodchickov equation}

The {\small KZ} differential equation 
\begin{equation}
\kappa \frac{\partial \Psi (z)}{\partial z_{i}}=H_{i}(z)\Psi (z),
\label{kz.1}
\end{equation}
appeared first as a \ holonomic system of differential equations on
conformal blocks in a {\small WZW} model of conformal field theory. Here $%
\Psi (z)$ is a function with values in the tensor product $V_{1}\otimes
\cdots \otimes V_{N}$ of representations of a simple Lie algebra, $\kappa
=k+g$ , where $k$ is the central charge of the model, and $g$ is the dual
Coxeter number of the simple Lie algebra.

One of the remarkable properties of the {\small KZ} system is that the
coefficient functions $H_{i}(z)$ commute and that the form $\omega
=\sum_{i}H_{i}(z)dz_{i}$ is closed \cite{RV}: 
\begin{equation}
\frac{\partial H_{j}}{\partial z_{i}}=\frac{\partial H_{i}}{\partial z_{j}}%
,\qquad \left[ H_{i},H_{j}\right] =0.  \label{kz.2}
\end{equation}

In this section we will identify $H_{i}$ with the $osp(2|1)$ Gaudin
Hamiltonians $G_{a}$ presented in the previous section and show that the
corresponding differential equations (\ref{kz.1}) can be solved via the
off-shell Bethe Ansatz method.

Let us now define the vector-valued function $\Psi (z_{1},...,z_{N})$
through multiple contour integrals of the Bethe vectors (\ref{off.2}) 
\begin{equation}
\Psi (z_{1},...,z_{N})=\oint \cdots \oint {\cal X}(u|z)\Phi
_{n}(u|z)du_{1}...du_{n},  \label{kz.3}
\end{equation}
where ${\cal X}$ $(u|z)={\cal X}$ $(u_{1},...,u_{n},z_{1},...,z_{N})$ is a
scalar function which in this stage is still undefined.

We assume that $\Psi (z_{1},...,z_{N})$ is a solution of the equations 
\begin{equation}
\kappa \frac{\partial \Psi (z_{1},...,z_{N})}{\partial z_{a}}=G_{a}\Psi
(z_{1},...,z_{N}),\quad a=1,2,...,N  \label{kz.4}
\end{equation}
where $G_{a}$ are the Gaudin Hamiltonians (\ref{gau.6}) and $\kappa $ is a
constant.

Substituting (\ref{kz.3}) into (\ref{kz.4}) we have 
\begin{equation}
\kappa \frac{\partial \Psi (z_{1},...,z_{N})}{\partial z_{a}}=\oint \left\{
\kappa \frac{\partial {\cal X}(u|z)}{\partial z_{a}}\Phi _{n}(u|z)+\kappa 
{\cal X}(u|z)\frac{\partial \Phi _{n}(u|z)}{\partial z_{a}}\right\} du,
\label{kz.5}
\end{equation}
where we are using a compact notation $\oint \left\{ \circ \right\} du=\oint
\ldots \oint \left\{ \circ \right\} $\ $du_{1}\cdots du_{n}.$

Using the Gaudin algebra (\ref{gau.17}) one can derive the following
non-trivial identity 
\begin{equation}
\frac{\partial \Phi _{n}}{\partial z_{a}}=\sum_{l=1}^{n}(-)^{l}\frac{%
\partial }{\partial u_{l}}\left( \frac{{\rm e}^{-u_{l}+z_{a}}\phi
_{(n-1)}^{l}}{\sinh (u_{l}-z_{a})}\right) ,  \label{kz.6}
\end{equation}
which allows us write (\ref{kz.5}) in the form 
\begin{eqnarray}
\kappa \frac{\partial \Psi }{\partial z_{a}} &=&\oint \left\{ \kappa \frac{%
\partial {\cal X}(u|z)}{\partial z_{a}}\Phi
_{n}(u|z)-\sum_{l=1}^{n}(-)^{l}\kappa \frac{\partial {\cal X}(u|z)}{\partial
u_{l}}\left( \frac{{\rm e}^{-u_{l}+za}\phi _{(n-1)}^{l}}{\sinh (u_{l}-z_{a})}%
\right) \right\} du  \nonumber \\
&&+\kappa \sum_{l=1}^{n}(-)^{l}\oint \frac{\partial }{\partial u_{l}}\left( 
{\cal X}(u|z)\frac{{\rm e}^{-u_{l}+za}\phi _{(n-1)}^{l}}{\sinh (u_{l}-z_{a})}%
\right) du.  \label{kz.7}
\end{eqnarray}
It is evident that the last term of (\ref{kz.7}) is vanishes, because the
contours are closed. Moreover, if the scalar function ${\cal X}(u|z)$
satisfies the following differential equations 
\begin{equation}
\kappa \frac{\partial {\cal X}(u|z)}{\partial z_{a}}=g_{a}{\cal X}%
(u|z),\qquad \kappa \frac{\partial {\cal X}(u|z)}{\partial u_{j}}%
=f_{j}^{(n-1)}{\cal X}(u|z),  \label{kz.8}
\end{equation}
we are recovering the off-shell Gaudin equation (\ref{off.11}) from the
first term in (\ref{kz.7}).

Taking into account the definition of the scalar functions \ $f_{j}^{(n-1)}$(%
\ref{off.7}) and $g_{a}$ (\ref{off.12}), we can see that the consistency
condition of the system (\ref{kz.8}) is insured by the zero curvature
conditions $\partial f_{j}^{(n-1)}/\partial z_{a}=\partial g_{a}/\partial
u_{j}$. Moreover, the solution of (\ref{kz.8}) is easily obtained 
\begin{equation}
{\cal X}(u|z)=\prod_{a<b}^{N}\sinh (z_{a}-z_{b})^{1/\kappa
}\prod_{j<k}^{n}\sinh (u_{j}-u_{k})^{1/\kappa
}\prod_{a=1}^{N}\prod_{j=1}^{n}\sinh (z_{a}-u_{j})^{-1/\kappa }.
\label{kz.9}
\end{equation}
This function determines the monodromy of $\Psi (z_{1},...,z_{N})$ as
solution of the trigonometric {\small KZ} equation (\ref{kz.4}) and these
results are in agreement with the Schechtman-Varchenko construction for
multiple contour integral as solutions of the {\small KZ} equation in an
arbitrary simple Lie algebra \cite{SV}.

\section{Highest Representations}

The generalization of our results for the highest representations of the $%
osp(2|1)$ algebra requests the knowledge of the corresponding algebraic
Bethe Ansatz which. It can be obtained from the Bethe Ansatz for the
fundamental representation using a fusion procedure. Nevertheless, we can
use the fact that the one-parameter operator families (\ref{gau.13}) form
the highest weight module of the infinite-dimensional Lie superalgebra. As
in the $sl_{2}$ case presented by Sklyanin in \cite{SKL1}, it is
characterized by the vacuum $\left| 0\right\rangle $%
\begin{equation}
{\cal H}(u|z)\left| 0\right\rangle =h(u|z)\left| 0\right\rangle ,\qquad 
{\cal V}^{+}(u|z)\left| 0\right\rangle =0,\qquad {\cal X}^{+}(u|z)\left|
0\right\rangle =0,  \label{high.1}
\end{equation}
the dual vacuum $\left\langle 0\right| $%
\begin{equation}
\left\langle 0\right| {\cal H}(u|z)=h(u|z)\left\langle 0\right| ,\qquad
\left\langle 0\right| {\cal V}^{-}(u|z)=0,\qquad \left\langle 0\right| {\cal %
X}^{-}(u|z)=0,\qquad \left\langle 0\right. |\left. 0\right\rangle =1,
\label{high.2}
\end{equation}
and the highest weight scalar function 
\begin{equation}
h(u|z)=\sum_{a=1}^{N}2s_{a}\coth (u-z_{a}),\qquad s_{a}=\frac{1}{2},1,\frac{3%
}{2},\cdots  \label{high.3}
\end{equation}

The solution of the Gaudin eigenvalue problem is given in term of Bethe
vectors $\Phi _{n}(u_{1},...,u_{n})$ given by (\ref{off.2}) with the
operators written in the $j$-representation (\ref{str.5}) and (\ref{str.7}).

It is convenient to express the Gaudin Hamiltonians $G_{a}$ in terms of a
generating function $t(u|z)$ which is obtained as the result of replacing $H$%
, $X^{\pm }$ and $V^{\pm }$ in the quadratic Casimir (\ref{str.2}) by ${\cal %
H}(u|z)$, ${\cal X}^{\pm }(u|z)$ and ${\cal V}^{\pm }(u|z)$, respectively: 
\begin{equation}
4t(u|z)=\sum_{a=1}^{N}\frac{{\cal G}_{a}(u)}{{\rm e}^{u-z_{a}}\sinh (u-z_{a})%
}+\sum_{a=1}^{N}\left( {\cal H}_{a}^{2}+\frac{s_{a}(s_{a}+1/2)}{\sinh
(u-z_{a})^{2}}\right) .  \label{high.4}
\end{equation}
where ${\cal G}_{a}(u)$ is given by (\ref{gau.6}).

Using a theorem \cite{GA2, SKL2} which states that $\Phi
_{n}(u_{1},...,u_{n})$ is an eigenvector of the commuting operators ${\cal G}%
_{a}(u)$ or, equivalently, of $t(u|z)$ if and only if the parameters $%
u_{1},...,u_{n}$ satisfy the Bethe equations 
\begin{equation}
\sum_{a=1}^{N}2s_{a}\coth (u-z_{a})=\sum_{k=1}^{n}\coth (u-u_{k}).
\label{high.5}
\end{equation}

The corresponding eigenvalue $\Lambda (u)$ of $4t(u|z)$ is then 
\begin{equation}
\Lambda (u)=\lambda ^{2}(u)-2\ \partial _{u}\lambda (u)  \label{high.6}
\end{equation}
where 
\begin{equation}
\lambda (u)=\sum_{a=1}^{N}2s_{a}\coth (u-z_{a})-\sum_{k=1}^{n}\coth (u-u_{k})
\label{high.7}
\end{equation}
From these results we conjecture that the off-shell Gaudin equation for the
highest representations is also given by (\ref{off.11}) with 
\begin{equation}
f_{j}^{(n-1)}=-\sum_{a=1}^{N}2s_{a}\coth (u_{j}-z_{a})+\sum_{k\neq
j}^{n}\coth (u_{j}-u_{k}),  \label{high.8}
\end{equation}
and 
\begin{equation}
g_{a}=\sum_{b\neq a}4s_{a}s_{b}\coth (z_{a}-z_{b})-\sum_{l=1}^{n}2s_{a}\coth
(z_{a}-u_{l}).  \label{high.9}
\end{equation}
Consequently, the monodromy of the function $\Psi (z_{1},...,z_{N})$ (\ref
{kz.3}) for the $j$-representation is given by 
\begin{equation}
{\cal X}(u|z)=\prod_{a<b}^{N}\sinh (z_{a}-z_{b})^{4s_{a}s_{b}/\kappa
}\prod_{j<k}^{n}\sinh (u_{j}-u_{k})^{1/\kappa
}\prod_{a=1}^{N}\prod_{j=1}^{n}\sinh (z_{a}-u_{j})^{-2s_{a}/\kappa },
\label{high.10}
\end{equation}
which also is in agreement with the Schechtman-Varchenko construction \cite
{SV}.

\section{Conclusion}

In this paper a graded $19$-vertex model was used to generalise previous
rational results connecting Gaudin magnet models and semi-classical
off-shell Bethe Ansatz of vertex models.

Using the semi-classical limit of the transfer matrix of the vertex model we
derive the trigonometric $osp(2|1)$ Gaudin Hamiltonians. The reduction of
the off-shell Gaudin equation \ to an eigenvalue equation gives us the exact
spectra and eigenvectors for these Gaudin magnets.  Data of the off-shell
Gaudin equation were used to show that a Jackson-type integral (\ref{kz.3})
is solution of the trigonometric {\small KZ} differential equation. \ 

In fact, this method had already been used with success to constructing
solutions of trigonometric {\small KZ} equations \cite{B2, CH} and elliptic 
{\small KZ}-Bernard equations \cite{B3}, for the six-vertex model and
eight-vertex model, respectively.

\vspace{1cm}{}

{\bf Acknowledgment:} This work was supported in part by Funda\c{c}\~{a}o de
Amparo \`{a} Pesquisa do Estado de S\~{a}o Paulo--FAPESP--Brasil, by
Conselho Nacional de Desenvol\-{}vimento--CNPq--Brasil and by Coordena\c{c}%
\~{a}o de Aperfei\c{c}oamento de Pessoal de N\'{\i}vel
Superior--CAPES-Brasil.

\end{document}